# Large scale chemical functionalization of graphene with nanometer resolution

Karolina Drogowska[1], Václav Valeš[1], Jan Plšek[1], Magdalena Michlová[1], Jana Vejpravová[2*], Martin Kalbáč[1*]

[1]J. Heyrovsky Institute of Physical Chemistry, v.v.i., Czech Academy of Sciences, Dolejškova 3, 18223 Prague 8, Czech Republic

[2]Department of Condensed Matter Physics, Faculty of Mathematics and Physics, Charles University, Ke Karlovu 5, 121 16 Prague 2, Czech Republic

**Abstract**

Anchoring various functional groups to graphene is the most versatile approach for tailoring its functional properties. To date, one must use a special tunneling microscope for attaching a molecule at a specific position on the graphene with resolution better than several hundred nanometers, however, achieving this resolution is impossible on a large scale. We demonstrate for the first time that chemical functionalization can be achieved with nanometer resolution by introducing strain with nanometer scale modulation into a graphene layer. The spatial distribution of the strain has been achieved by transferring a single-layer graphene (SLG) onto a substrate decorated by a few nm large nanoparticles (NPs). By changing the number of NPs on the substrate, the amount of locally strained SLG increases, as confirmed by atomic force microscopy (AFM) and Raman spectroscopy investigations. We further carried out hydrogenation and fluorination on the SLG with increasing amount of nanoscale corrugations. Raman spectroscopy, AFM and X-ray photoelectron spectroscopy revealed unambiguously that the level of functionalization increases proportionally with the number of NPs, which means an increasing number of the locally strained SLG. Our approach thus enables control of the amount and the position of functional groups on graphene with nanometer resolution.

[*] Corresponding authors. Tel.: +420 951 552 735. Email: jana@mag.mff.cuni.cz (Jana Vejpravová). Tel.: +420 266 053 804. Email: martin.kalbac@jh-inst.cas.cz (Martin Kalbáč)

1. Introduction

Chemical functionalization of graphene is a versatile tool for tuning its intrinsic properties and for controlling the interaction of graphene with the environment, giving rise to many remarkable applications [1–6]. In particular, tuning the band structure from a semimetal to a semiconductor and/or an insulator is the evergreen in graphene research. One of the key strategies of SLG modification is chemical functionalization, in particular hydrogenation and/or fluorination [7–14]. Moreover, the prior hydrogenation of pristine graphene creates the possibility for its secondary functionalization with chemical species that do not react directly with the pristine graphene [15]. Similarly, fluorinated graphene is also a versatile platform that enables comfortable attachment of various functional groups onto SLG [6].

Obviously, the properties of chemically functionalized graphene strongly depend on the degree of its functionalization [4,16,17]. Thus, control over graphene reactivity, in particular for SLG on substrates, is a key for further advancements of its applications. The advantage of SLG on substrate, if the SLG has been mechanically exfoliated or grown by chemical vapor deposition (CVD) rather than chemically exfoliated, is that the substrate can be used for fine and local tuning of SLG properties. So far, several strategies have been reported for control of chemical functionalization. For example, reactivity of SLG on copper can be influenced by changing the surface phase orientation of the copper [18]. Other approaches such as heat treatment [19] and creation of defects prior to functionalization have been tested [20,21].

The real control over reactivity by means of number of functional groups and their position on the graphene remains a major challenge, especially for larger scale areas. Although some scanning tunneling microscopic techniques can achieve local functionalization with nanometer resolution, this approach is obviously not relevant for any practical application. It has also been demonstrated that diazonium chemistry on SLG is driven by graphene/substrate interaction via doping, and a spatial resolution of tens to hundreds of micrometers can be achieved by using reactivity imprint lithography [22]. As reported by Goler et al. [23] and Tozzini et al. [24], the reactivity of graphene may be increased via manipulation of graphene curvature. This strategy has been applied very recently by Odom and co-workers, who have achieved spatially selective graphene functionalization with a spatial resolution of a few hundred nanometers using multiscale wrinkles [25].

Nevertheless, there is still an open question about the dominance of doping and strain making CVD graphene on substrates prone to anchoring functional groups via different reaction routes.

In order to disentangle the role of the two key driving forces for on-graphene chemistry, one has to achieve spatially resolved modulation of the doping and strain.

We have demonstrated previously [26,27] that the density of wrinkles in CVD-grown SLG on a $SiO_2$/Si substrate can be controlled by decorating the underlying substrate with uniform nanoparticles (NP). Thus, one can achieve a specific landscape on the SLG, which is a superposition of a flat graphene (adhered to the substrate) and a corrugated graphene (delaminated wrinkles). The wrinkle density can be tuned easily by modulating NP concentration, and consequently the sculptured SLG reveals spatial modulation of the doping and the strain on a nanometer scale, which can be addressed using Raman spectroscopy.

In this study, we demonstrate that, through local control of SLG topography using substrates decorated with a different number of uniform NPs, we can systematically tune the level and the resolution of chemical functionalization of a large area of CVD-grown SLG on $SiO_2$/Si substrate. In order to probe the reactivity of the modulated SLG, we tested two principal methods: 1. hydrogenation via annealing of SLG in a high-pressure, molecular hydrogen atmosphere, and 2. fluorination performed by exposing the SLG to $XeF_2$ in the gas phase.

The choice of the two approaches for covalent functionalization enabled us to inspect the process at low and high levels of functionalization, respectively. The proposed hydrogenation procedure is not considered a very efficient method of graphene hydrogenation and does not result in high conversion [15]. Therefore, it enables monitoring tiny changes in reactivity of SLG as a function of wrinkle density. On the other hand, fluorination is a very efficient way to create covalently functionalized SLG and it enables researchers to understand reactivity at a higher level of functionalization.

We demonstrate here that the level of hydrogenation and fluorination of CVD-grown SLG increases proportionally with the number of NPs on the underlying $SiO_2$/Si substrate, which means an increasing number of nanoscale corrugations (delaminated wrinkles). By applying a correlation analysis of large sets of Raman spectra, we discovered that in this case the key factor influencing reactivity is a mechanical strain accommodated in the locally curved SLG due to nano-modulated topography. Our strategy offers new insight into the on-surface chemistry of graphene deposited on substrates and opens the door for highly controllable tailoring of the physio-chemical properties of CVD graphene and other two-dimensional materials with nanometer resolution on large scale.

## 2. Experimental

2.1 Sample preparation

Graphene was grown using the standard CVD method, as described elsewhere [28]. Briefly, polycrystalline copper foil (0.025 mm thick) was annealed at a temperature of 1000°C in a hydrogen flow of 50 sccm for 20 minutes, at which time the methane precursor was switched on. Graphene growth was carried on for 30 minutes with the flow of methane at 1 sccm. Then the sample was annealed another five minutes in $H_2$ to etch the top layer and cooled down to room temperature. The pressure during the whole process was kept at about 0.35 Torr. The graphene was then transferred to a $Si/SiO_2$ substrate decorated with NPs using nitrocellulose-based transfer, as described by Hallam et al. [29].

Nanoparticles (cobalt ferrite, $CoFe_2O_4$) were synthesized by a hydrothermal procedure described in a previous work [30]. In a general procedure, 10 mmol (0.400 g) of NaOH were dissolved in 2 mL of water. Then, 10 mL of ethanol and 12 mmol (3.39 g) of oleic acid were added with stirring to create a transparent sodium oleate solution. 2 mmol (0.808 g) of iron nitrate and 1 mmol (0.291 g) of cobalt nitrate were dissolved in 20 mL of distilled water and added to the mixture with vigorous stirring. The product was transferred to an autoclave tube (Berghof DAB-2 autoclave with 40 mL teflon liner). The autoclave was closed and placed into a preheated oven at 200°C for 11 h. After cooling, the phase that contained nanoparticles was separated and cleaned four times by re-dispersion in 10 mL of hexane and precipitation by 15 mL of ethanol. For more efficient re-dispersion, sonication was used. Finally, particles were separated by a permanent magnet. After washing, nanoparticles were re-dispersed into 10 mL of hexane and dispersion was centrifuged at 4500 rpm for five minutes to remove larger aggregates. A typical TEM image of the final product is shown in Figure S1. The NP were deposited on the substrates using spincoating method.

2.2 Chemical functionalization

Hydrogenation of graphene was carried out in the high-pressure autoclave, a Berghof HR-100. Prior to the hydrogenation process, the chamber was flushed with hydrogen (Messer, 6.0) to remove the rest of the air. Then, the chamber was filled with hydrogen gas under 7 bars of pressure and the temperature was increased to 200°C. The hydrogenation process was carried on for two hours. Details may be found in a study by Drogowska et al. [15].

Fluorination of graphene was performed in a homebuilt apparatus by exposing the samples to $XeF_2$ in the gas phase. The sample was placed in a vacuum chamber that was isolated from the pump after obtaining a pressure of $2·10^{-4}$ mbar. The vacuum chamber was then connected to the solid $XeFe_2$ (Aldrich, 99.99%) reservoir. Fluorination was carried out for one second.

2.3 Microscopic and spectroscopic characterization

AFM images were captured using the Dimension Icon atomic force microscope (Bruker, Inc.). The RFESP-75 probes ($k$ = 3 N/m, $f_0$ = 75 kHz, L = 225 nm; Bruker, Inc.) were used in tapping mode with 1024 lines of resolution and a 0.6 Hz scan rate. The amplitude setpoint was chosen in order to minimize the forces acting on the graphene layer. The maximum size of the captured images was 5x5 $\mu m^2$. Gwyddion software was used to process the images, applying line-by-line, first-order levelling and scar removal. Consequently, grain analysis was used to evaluate surface coverage of the samples by the nanoparticles and to determine the extent of corrugation of the graphene layer by the NPs.

Raman spectroscopy was performed using a WITec alpha300 R spectrometer equipped with a piezo stage with a 532 nm excitation wavelength. The laser power was kept below 1 mW to avoid sample overheating. The laser was focused on the sample with a 100× objective to a spot with a diameter of around 500 nm. The Raman maps of 30×30 $\mu m^2$ (typically containing 900 spectra) were acquired with every lateral step of 1 µm in both directions. The data were analyzed using a homemade routine in Octave, enabling automated fitting of the large data sets considering the peaks in the form of pseudo-Voigt functions.

The XPS measurements were performed in a VG ESCA3 Mk II electron spectrometer. Al K$\alpha$ radiation was used for excitation of the electrons. The photoemitted electrons were energy analyzed using a hemispherical analyzer operated at a constant pass energy of 20 eV.

3. Results and discussion

3.1 Preparation of SLG with spatially modulated doping and strain

First, we studied two series of modulated SLG samples on Si/SiO$_2$ substrates decorated with NPs. For each series, we prepared three different hexane dispersions of the NPs: low concentration (two samples termed H1 and F1), high concentration (samples termed H2 and F2), and dispersion with larger objects - aggregates (sample H3 subjected to hydrogenation).

A reference sample without NPs (F0) was also prepared. The samples labelled "H" were designated for hydrogenation, whereas those labelled "F" were designated for fluorination. Typical AFM topography images for the hydrogenation series are shown in Figure 1. The images for the fluorination series are given in Figure S2. The top row of Figure 1, (a) – (c), corresponds to the substrates decorated with NPs before deposition of SLG, and the bottom row, (d) – (f), represents the AFM topography of the H1, H2, and H3 samples, respectively. A 3D view onto a SLG transferred onto a sample with low (H1) and high (H3) concentration of NPs is given in panels (j) and (k), respectively. Each sample reveals homogenous coverage of the substrate with NPs. The density of the NPs coverage was found to be 23, 128, and 231 NPs per $\mu m^2$ for the H1, H2, and H3 samples, respectively.

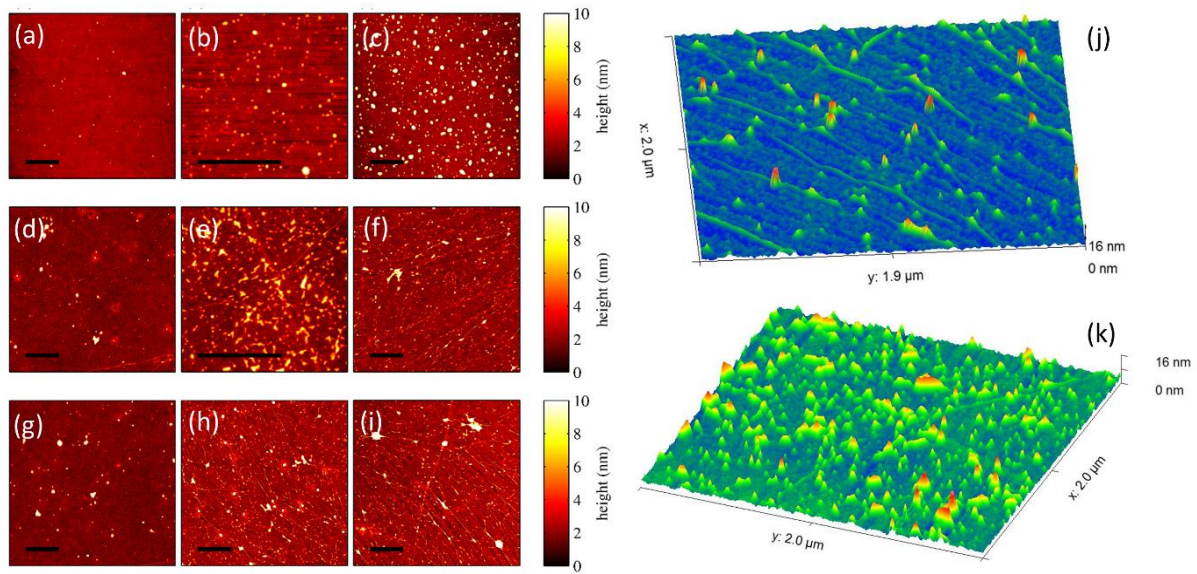

*Figure 1. AFM images of the Si/SiO$_2$ substrates decorated with different concentrations of NPs: 23(a), 128(d) and 231 (c) $\mu m^{-2}$ (samples H1, H2 and H3). Panels (d-f) display the AFM images of the CVD-grown graphene deposited on the substrates presented in the panels (a-c). Panels (g-i) display topography of the graphene shown in panels (d-f) after hydrogenation. The scale bar is set to 1 µm. Panels (j) and (k) correspond to a typical landscape of the graphene layer transferred on the substrate decorated with NPs with low and high concentration, respectively.*

Analysis of the AFM data performed on the samples, which were in the next step hydrogenated (H1, H2, and H3), did not reveal any significant differences in NP concentration or homogeneity before or after transferring the SLG. Therefore, the SLG transfer did not cause any harm to the NP-decorated substrate. Conforming to expectations, the deposition of the SLG onto substrates decorated with NPs resulted in the formation of a specific landscape on the SLG composed of flat and delaminated fractions, as shown in the Figure 1 (d – e).

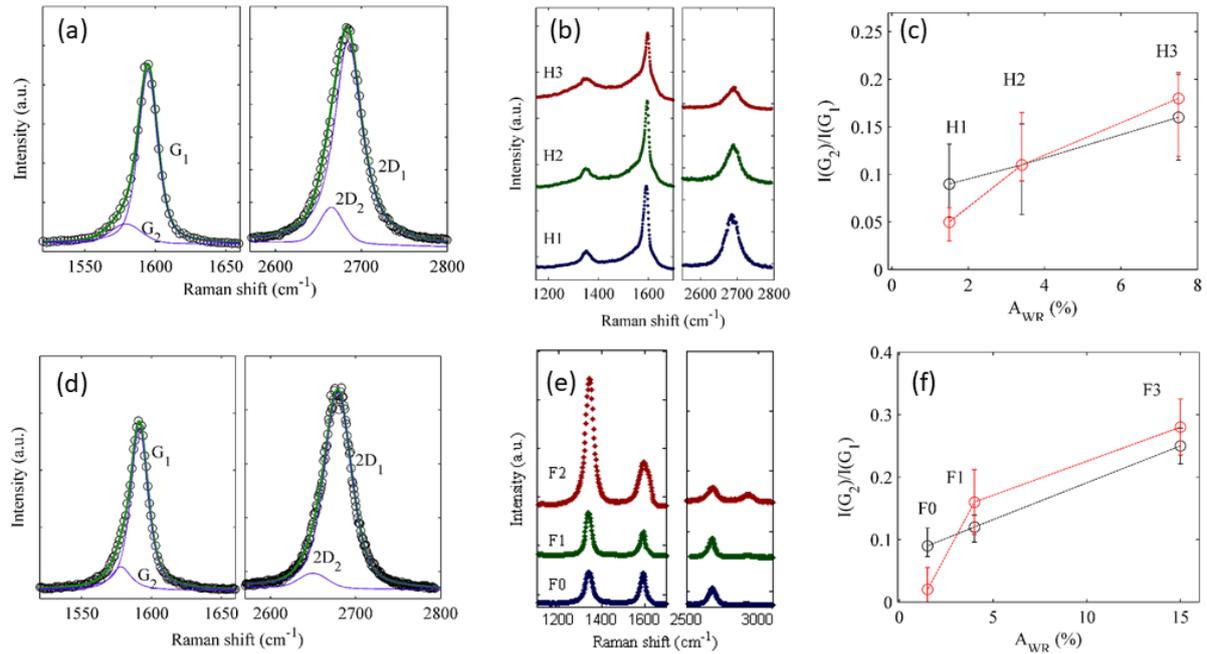

*Figure 2. Representative Raman spectra before (a, d) and after (b, e) chemical functionalization and correlation of the $G_2/G_1$ integral intensity ratio to the wrinkle density given by the relative delamination area (c, f). (a) Typical Raman spectrum of graphene deposited on the Si/SiO$_2$ substrate with the wrinkle density of 7.5 % (sample H3, concentration of NPs: 231 $\mu m^{-2}$). The black dots represent the experimental data, blue curves correspond to the $G_1$, $G_2$ and $2D_1$, $2D_2$ modes, whereas the green line represents their superposition. (b) Raman spectra of the H1, H2, and H3 samples after hydrogenation. (c) Dependence of the $I(G_2)/I(G_1)$ ratio on the relative delamination area, which is directly related to the number of wrinkles and scales with the concentration of NPs. The black and red circles represent the data before and after the hydrogenation process, respectively. (d) Typical Raman spectrum of graphene deposited on the Si/SiO$_2$ substrate with a wrinkle density of 3.5 % (sample F1, concentration of NPs: 231 $\mu m^{-2}$) with its fit prior to fluorination. The black dots represent the experimental data, blue curves correspond to both contributions of the G mode, $G_1$ and $G_2$, whereas the green line represents their superposition. (e) Raman spectra of the F0, F1, and F2 samples after fluorination. (f) Dependence of the $I(G_2)/I(G_1)$ ratio on the delamination area. The black points represent data prior to the fluorination process, whereas the red points represent data after the fluorination process.*

We further subjected the AFM data to topography analysis, which revealed height histograms of the delaminated wrinkles for all samples, as shown in Figure S3. According to the histograms, the mean height of the wrinkles in the modulated samples was about 3–4 nm, compared to the reference sample F0 with negligible wrinkle density (2 nm). We can conclude that the transfer process introduces a minor number of wrinkles with 25–50 % height reduction compared to the SLG transferred onto NP-decorated substrates. The level of delamination represented by the relative delaminated area, $A_{WR}$, was found to be 1.5 %, 3.4 %, and 7.5 % for the H1, H2, and H3 samples, respectively.

The AFM analysis also confirmed homogenous distribution of the NPs on the substrate of the F-series, and their concentration was found to be 36 and 191 NPs per $\mu m^2$ for the F1 and F2 samples, respectively. Deposition of SLG onto those substrates resulted in a wrinkle density of 3.5 % and 14.5 %, respectively, while the wrinkle density of the reference F0 sample was found to be below 1.5 %, which is comparable to the results obtained for the H1 sample.

We further performed Raman spectroscopy by mapping each sample at an area of about 30 x 30 $\mu m^2$. We detected all typical features of SLG: G and 2D modes. (Note: Prior to functionalization, the Raman spectra did not reveal the presence of the D mode, showing that the samples were almost defect-free after SLG transfer to the NP-decorated substrates.)

The G mode corresponding to the $E_{2g}$ phonon at the Brillouin zone center was observed at a Raman shift of about 1590 cm$^{-1}$, suggesting doping because of the Si/SiO$_2$ substrate [31]. The G mode should be theoretically modelled by a single Lorentzian peak. However, this approximation is not valid in real samples as variation of the doping (and strain) in the area of the laser spot can cause distribution of peak parameters. To account for this variation, the G mode is better approximated by a pseudo-Voigt function. Nevertheless, in case of a graphene with modulated topography, the G mode can be decomposed into two peaks, $G_1$ and $G_2$ [26,32]. While the $G_1$ mode accounts for the flat SLG adhered to the substrate, the $G_2$ mode can be treated as a fingerprint of the delaminated fraction of the SLG due to topographic corrugations [26,32]. Thus, in our procedure, we evaluated the G mode as a superposition of the $G_1$ and $G_2$ modes.

The Raman shift of the 2D mode originating from the second-order phonon process was found to be about 2680 cm$^{-1}$, pointing to insignificant strain in the SLG [33]. Please note that we also describe the 2D mode with two components, $2D_1$ and $2D_2$, where $2D_1$ corresponds to the flat and $2D_2$ to the delaminated graphene, respectively. Typical spectra of the samples with the highest level of delamination (H3 and F2), together with the corresponding fits, are shown in Figure 2, panels (a) and (d), respectively. Spatial maps of the Raman shifts and intensities for all modes of the H2, H3, F1, and F2 samples are given in SI, Figures S4 and S5.

To inspect the topography fingerprints in the Raman spectra reflecting the spatial modulation of the strain and doping, we performed decomposition of the Raman spectral parameters in accordance with Lee et al. [33]. A typical correlation diagram can be constructed by plotting the pairs of the Raman shift of the G and 2D, as the Raman shifts of G ($\omega_G$) and 2D ($\omega_{2D}$) are sensitive to both the doping and strain, but with very different fractional variations, ($\Delta\omega_{2D}/\Delta\omega_G$). The phase space can be decomposed into a series of doping and strain lines intersecting at point $P_0$ (1582, 2637) cm$^{-1}$, corresponding to a pristine SLG. The slope of the biaxial strain line was reported in the range of 2.25–2.8 [34,35], and the slope of the doping line was about 0.7 [36]. For the biaxial strain approximation, we can assess the magnitude and sign of the strain within the doping line, considering the $\Delta\omega_{2D}$ is about -144 cm$^{-1}$ for a 1% strain [33]. The doping can be estimated from the $\Delta\omega_G$ value within the strain line using the formula proposed by Das et al. [37], assuming hole doping to be due to the presence of the $SiO_2$ substrate. Each point in the correlation diagram can thus be described as a linear combination of the unit vectors corresponding to the strain, **e**$_\varepsilon$, and doping, **e**$_n$, with the origin of $P_0$ (for details, see SI, Figure S6).

Figure 3 represents correlation diagrams for all samples prior to (color plots) and after (grey plots) chemical functionalization. The $G_1$ - $2D_1$ pairs corresponding to the flat (adhered) graphene are located at the zero strain line, while the points are clearly extended along a doping line ($n \sim 5.10^{12} – 10^{13}$ cm$^{-2}$), suggesting dominant influence of doping on the Raman shift of the correlated modes. In contrast, the $G_2$ - $2D_2$ pairs corresponding to the delaminated area are located in the low doping region, with moderate compressive strain not exceeding 0.15 %. The results are in agreement with previous studies that suggest higher doping of adhered flat graphene due to strong interaction with the underlying $SiO_2$/Si substrate and low doping accompanied by moderate strain in the delaminated wrinkles [26,38,39].

As suggested above, the intensity of the $G_2$ mode increases linearly with the delamination level of the SLG [26,27]. Therefore, the ratio of the $G_2$ and $G_1$ intensities, $I(G_2)/I(G_1)$, serves as the measure of the degree of SLG delamination. The calculated median values of $I(G_2)/I(G_1)$ for the H and F series are shown in Figure 2, (c) and (f), respectively. The results clearly show monotonous increase of $I(G_2)/I(G_1)$, suggesting an increase in the $A_{WR}$ due to the increasing number of delaminated wrinkles in correspondence with the AFM topography. The obtained results fit to the general scenario as confirmed by plotting $I(G_2)/I(G_1)$ versus $A_{WR}$ (Figure S7) for the current data and the data acquired in the previous study [26]. Our approach for quantification of the topographic corrugations in graphene is also an excellent tool for addressing chemical functionalization, as discussed below.

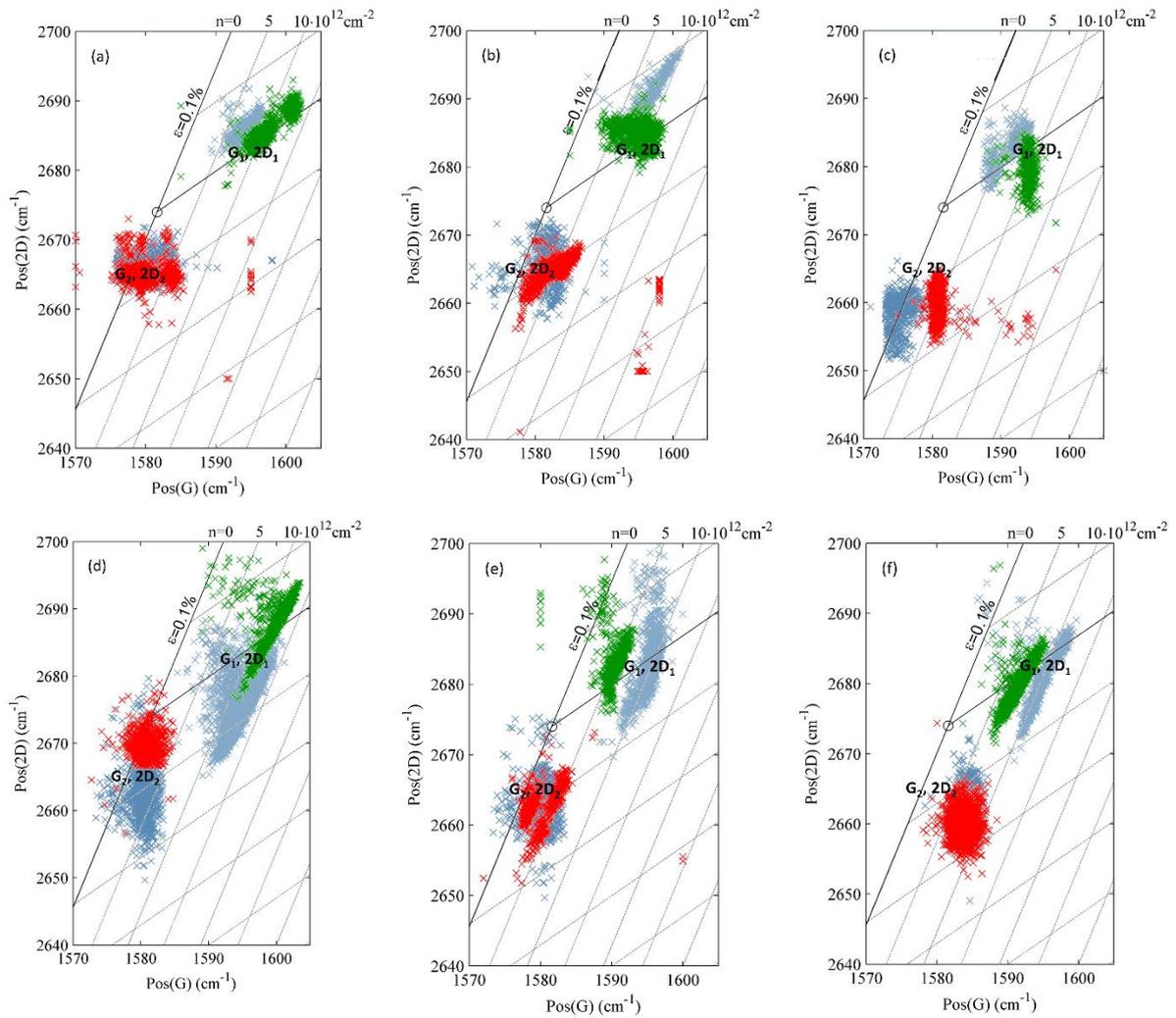

*Figure 3. G – 2D correlation diagrams for the pristine and hydrogenated (a – c)/fluorinated (d – f) samples constructed from the fits of the Raman maps. Panels (a – c) correspond to the H1, H2, and H3 samples and panels (d – f) correspond to the F0, F1, and F2 samples, respectively. The green and red points correspond to the $G_1$, $2D_1$ (adhered, flat fraction) and $G_2$, $2D_2$ (delaminated, wrinkled fraction) correlation pairs. The grey points correspond to the data points obtained after hydrogenation (a – c)/fluorination (d – f). The grids represent doping and strain lines.*

## 3.2 Chemical functionalization of the SLG with modulated topography

We further subjected the samples with different numbers of delaminated wrinkles to hydrogenation and fluorination. In order to check for possible damage to the samples during exposure to hydrogen or fluorine, routine AFM characterization was performed after hydrogenation (Figure 1 (f) – (h)) and fluorination (Figure S1 (f) – (h)). The inspection did not show any significant changes in NP concentration and homogeneity, or in the SLG topography. The samples were further subjected to Raman spectroscopy investigation. Typical spectra of hydrogenated samples are shown in Figure 2 (b). We observed that the hydrogenation results in the D mode in the spectra, which is in general viewed as an indirect proof of chemical functionalization, related to a change of the hybridization of carbon atoms from $sp^2$ to $sp^3$. Furthermore, it is obvious that the intensity of the D mode is directly related to the number of delaminated wrinkles represented by the $A_{WR}$ and also directly linked to the intensity of the $G_2$ mode. Moreover, the intensity of the 2D mode decreased rapidly after hydrogenation. Therefore, there is a clear dependence of the increase of the D mode and decrease of the 2D mode on $A_{WR}$ (number of topographic corrugations).

In order to quantify the enhancement of SLG reactivity with hydrogen, we have calculated the median values of the ratio of the intensities of the D and G modes, I(D)/I(G). As shown in Figure 4(a), hydrogenation of the SLG led to an increase in D mode intensity proportional to the $A_{WR}$ (number of wrinkles), which is especially emphasized for the graphene deposited on the NPs with the highest concentration (sample H3). The linear increase of the D mode as a measure of the level of hydrogenation thus clearly correlates with an increasing number of topographic corrugations in the modulated SLGs.

It is also worth mentioning that, after hydrogenation, the values of $I(G_2)/I(G_1)$ are slightly higher, especially for the SLG with high $A_{WR}$. This result can be explained by an overall increase in SLG delamination caused by the attachment of hydrogen atoms to the graphene lattice [40]. This hypothesis is somewhat supported by the shift of the $G_1 - 2D_1$ points in the G – 2D correlation diagrams towards low doping. A similar shift is clearly observed for the $G_2 - 2D_2$ correlation pairs of the H3 sample only, which move towards zero doping line.

We have further performed the same experiment, i.e. functionalization of SLG deposited on substrates decorated with NPs, using fluorination. In contrast to hydrogenation, this procedure is known as a very efficient reaction on graphene.[41] Figure 2(d) shows a typical Raman spectrum of the fluorinated sample with the highest $A_{WR}$ value (F2). As in the case of the H3 sample, the spectrum reveals features typical for delaminated graphene, in particular the asymmetric shape of the G and the 2D modes, which can be further decomposed into the $G_1$ and $G_2$ and the $2D_1$ and $2D_2$ components, respectively.

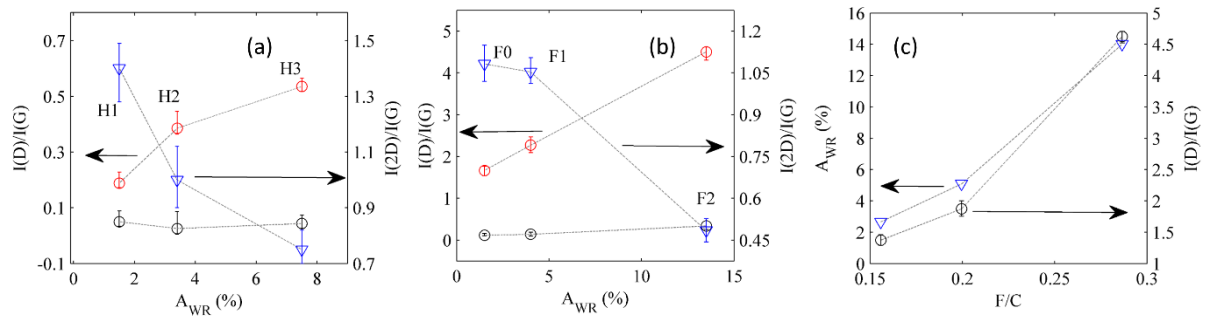

*Figure 4. Correlation of wrinkle density given by the relative delaminated area to the level of chemical functionalization determined from the I(D)/I(G) ratio and XPS. Dependence of the I(D)/I(G) and I(2D)/I(G) ratios on the relative delamination area prior to and after hydrogenation (a) and fluorination (b). The values represent median values obtained from the Raman map, whereas the error bars are the first and third quartiles. The black and red circles represent the results obtained from the samples before and after functionalization, respectively. Panel (c) demonstrates the excellent match between the I(D)/I(G) ratio (black circles), representing the level of fluorination, and the relative fluorine to carbon ratio (F/C) obtained from XPS (blue triangles).*

Figure 2(f) shows a clear linear dependence of $I(G_2)/I(G_1)$ versus $A_{WR}$, which is the measure of SLG corrugation. Fluorination resulted in a slight increase of the $I(G_2)/I(G_1)$ ratio as compared to the untreated samples, and as observed in the case of the hydrogenated samples. The G – 2D correlation diagrams also remain very similar to those of the F0, F1, and F2 samples prior to fluorination. However, there is a clear tendency for the $G_1 – 2D_1$ clouds to shift towards higher doping, whereas the $G_2 – 2D_2$ pairs show only a moderate change in the strain. Please note that the behavior of the adhered SLG represented by the $G_1 – 2D_1$ clearly contrasts with that of the hydrogenated samples, where the $G_1 – 2D_1$ points moved to the low doping region. This effect can tentatively be explained by significant charge transfer from the fluorine to the SLG, typically observed for graphene fluorinated by the same protocol [41].

We finally explored the effect of the fluorination carried out on SLG with different $A_{WR}$ values. Figure 2 (e) displays the Raman spectra for the fluorinated SLG deposited on the substrates decorated with different concentrations of NPs, giving rise to different $A_{WR}$ (F1 and F2). Figure 2 also displays the Raman spectra for the reference sample (F0). The functionalization with fluorine is evidenced by the appearance of the D and D' modes, as well as by rapid suppression of the 2D mode, especially pronounced for the strongly delaminated SLG (F2).

The median values of the ratio of the intensities of the D and G modes, I(D)/I(G), representing the measure of the degree of functionalization, is shown in Figure 4(b). The curve follows a linear dependence on the $A_{WR}$, suggesting an increasing degree of fluorination with an increasing number of delaminated wrinkles. These results confirm that the enhancement of SLG reactivity occurs due to the increasing fraction of the curved SLG surface, as discussed previously for the hydrogenated SLG. Consequently, we conclude that the local strain is most likely the governing factor for enhancement of the gas phase hydrogenation and fluorination of SLG on $SiO_2$/Si substrate.

In order to verify the results obtained by Raman spectroscopy, which suggests the enhancement of the chemical activity of SLG, we carried out XPS analysis of the fluorinated samples (as hydrogen cannot be detected by XPS). XPS serves as direct evidence of anchoring fluorine onto the modulated SLGs. The fluorine content was calculated from the C1s and the F1s XPS spectra for the samples F0, F1, and F2 (shown in SI, Figure S8). Because all of the samples were fluorinated under the same protocol, they were dosed with the same amount of $XeF_2$. (If there is no effect on the substrate, the amount of fluorine should be the same for all samples.) However, Figure 4(c) clearly shows a linearly increasing F/C ratio with wrinkle density (represented by $A_{WR}$). It is thus obvious that an increase of NP concentration on the substrate causes an increase of the number of wrinkles (curved surface fraction represented by $A_{WR}$), which resulted in higher reactivity of the SLG at the corrugated sites and an increase of the fluorine content.

To encapsulate the proposed scenario, we have finally plotted the dependence of the I(D)/I(G) ratio determined from the Raman correlation analysis on the F/C content obtained from XPS. The results presented in Figure 4(c) clearly demonstrate a linear proportionality of the I(D)/I(G) ratio, the common measure of the functionalization degree. The outcomes obtained from the two independent techniques are in a perfect match, and they explicitly prove that, through the manipulation of SLG curvature (i.e., through controlled transfer of the SLG onto NP-decorated substrates), one can achieve control of the number of functional groups on SLG.

We would like to point out that because of the very local character of the topographic corrugation in our samples, we revealed that the approach is powerful because of spatial localization of the anchored species. Literally speaking, the chemical reactions take place preferentially at the position of the delaminated wrinkle. Therefore, we are able to achieve spatially modulated chemical functionalization with a resolution of a few nanometers on macroscopic sample scales because the limitation is given by the size of the transferred SLG flake, which can easily range from a few hundred micrometers to many centimeters nowadays.

## 4. Conclusions

We investigated how SLGs with different numbers of topographic corrugations are prone to chemical functionalization. We developed a robust procedure for introducing a defined number of corrugations (delaminated wrinkles) into the SLG. We transferred the SLG onto $SiO_2$/Si substrates decorated by a different number of uniform NPs. We selected two different protocols of gas phase functionalization: hydrogenation via annealing in a hydrogen atmosphere and fluorination via reaction with $XeF_2$, the latter being much more efficient than the former. Using a Raman correlation breakdown corroborated with AFM topography analysis, we succeeded in quantifying of the number of wrinkles (delaminated SLG) and assigning the level of doping and strain in the unperturbed and delaminated SLG fractions. We clearly demonstrated that, with an increasing number of corrugations, the extent of both hydrogenation and fluorination of the sample increases. We finally completed the puzzle with XPS measurements, which confirmed unequivocally that the enhancement of reactivity giving rise to increasing fluorine content occurs due to the increasing amount of curved SLG accommodated in the delaminated wrinkles. Our results suggest that the key factor for enhancement reactivity of SLG is the strain imprinted in the local curvature of the SLG, making the graphene π orbitals more susceptible to accept other species. Regarding future prospects, our work is a keen plug in the on-surface graphene chemistry socket, unravelling novel strategies for 2D materials engineering at the nanoscale. For example, we can easily achieve semimetal – insulator architectures [25] with nanometer modulation or even diamagnetic – para/ferromagnetic [42] networks, giving rise to novel graphene-based devices with countless applications.

## Acknowledgements

This research was funded by the European Research Council (ERC-Stg-716265), Czech Science Foundation (18-20357S) and Ministry of Education, Youth and Sports of the Czech Republic under Operational Programme Research, Development and Education, project Carbon allotropes with rationalized nanointerfaces and nanolinks for environmental and biomedical applications (CARAT), number CZ.02.1.01/0.0/0.0/16_026/0008382. The authors acknowledge the assistance provided by the Research Infrastructures NanoEnviCz (Project No.

LM2015073) supported by the Ministry of Education, Youth and Sports of the Czech Republic, and the project Pro-NanoEnviCz (Reg. No. CZ.02.1.01/0.0/0.0/16_013/0001821) supported by the Ministry of Education, Youth and Sports of the Czech Republic, and the European Union - European Structural and Investments Funds in the frame of Operational Programme Research Development and Education.